\documentstyle[aps,graphicx,epsf,floats,goodfloat]{revtex}

\begin{document}

\draft

\twocolumn[\hsize\textwidth\columnwidth\hsize\csname@twocolumnfalse%
\endcsname

\title{Lower Critical Dimension of Ising Spin Glasses
}

\author{A.~K.~Hartmann\cite{alex:email}
and 
A.~P.~Young\cite{peter:email}
}
\address{Department of Physics, 
University of California, Santa Cruz CA 95064, USA}

\date{\today}
\maketitle
\begin{abstract}
Exact ground states of two-dimensional Ising spin glasses with Gaussian
and bimodal ($\pm J$) distributions of the disorder are calculated using a
``matching''algorithm, which allows 
large system
sizes of up to $N=480^2$ spins to be investigated. We study domain walls induced by
two rather different types of boundary-condition changes,
and, in each case, analyze the system-size dependence of 
an appropriately defined 
``defect energy'', which we
denote by $\Delta E$.
For Gaussian disorder, we find a power-law behavior 
$\Delta E \sim L^{\theta}$, with
$\theta=-0.266(2)$ and $\theta=-0.282(2)$ for the two types of
boundary condition changes. These results are in reasonable agreement with each
other, allowing for small systematic effects. 
They also agree well with earlier
work on smaller sizes. The
negative value indicates that
two dimensions is below the
lower critical dimension $d_c$.
For the $\pm J$
model, we obtain a {\em different}\/ result, namely
the domain-wall energy saturates at a nonzero value for $L\to \infty$,
so $\theta = 0$,
indicating that the lower critical dimension for the $\pm J$ model
{\em exactly}\/ $d_c=2$.
\end{abstract}

]

Although most of the effort to understand spin
glasses\cite{revievSG} has been concerned with
the behavior in three dimensions, the situation in two dimensions is also not
fully understood.
For a Gaussian distribution of disorder,
it is widely accepted that\cite{mcmillan1984,bray84,rieger1996},
$T_c=0$ and 
two--dimensions is {\em below}\/ the lower critical dimension $d_c$.
However, for bimodal
($\pm J$) disorder, inconsistent results have been found. Using exact
ground-state calculations of small systems ($N\le 30^2$)
and introducing domain walls in the system 
by changing the boundary conditions,  Kawashima and Rieger\cite{kawashima1997}
also concluded that $T_c =0$ and probably $d_c >2$, though they 
could not rule out marginal behavior, i.e.  $d_c=2$.
On the other hand, from  Monte-Carlo (MC)
simulations\cite{shirakura1997} on sizes $N\le 18^2$, it was claimed that
$T_c>0$.
A calculation of domain-wall energies using
a cluster MC method\cite{matsubara1998} for sizes $N\le 24^2$ 
indicated marginal behavior (i.e. $d_c = 2)$, and this was interpreted as
evidence for the finite-$T$ transition claimed in
Ref.~\onlinecite{shirakura1997}.
Recently, Houdayer\cite{houdayer2001} used a
more efficient cluster MC algorithm,
which allows larger systems ($N\le 100^2$)
to be studied down to low temperature $T=0.1$.
Houdayer finds that $T_c=0$ but with an
exponential divergence of the correlation length for $T\to 0$,
which is consistent with $d_c=2$, in agreement with earlier work by Saul and
Kardar\cite{saul1993}.

To try to clarify the situation, we calculate here
ground-state domain-wall energies $\Delta E$ 
using an {\em exact} polynomial-time matching
algorithm\cite{bieche1980} for much larger system sizes
($N \le 480^2$)
than in previous work.
For the Gaussian model
our results are consistent with earlier calculations and
we find
that $T_c=0$ and $d_c > 2$.
For the $\pm J$ model, we find marginal behavior expected at the lower
critical dimension, and so conclude that $d_c = 2$.

The model consists  of $N=L^2$ Ising spins $S_i=\pm 1$ on a square
lattice with the Hamiltonian
\begin{equation}
{\cal H} = -\sum_{\langle i,j \rangle} J_{ij} S_iS_j\,,
\end{equation}
where the sum runs over all pairs of nearest neighbors $\langle i,j \rangle$
and the $J_{ij}$ are quenched random variables. Here, we consider
two kinds of disorder
distributions: (i) Gaussian with zero mean and variance one,
and
(ii) a bimodal distribution $J_{ij}=\pm 1$ with equal probability.

In greater than two dimensions, or in the presence of a magnetic field,
the exact calculation of spin-glass ground states belongs
to the class of NP-hard problems\cite{barahona1982,opt-phys2001}. 
This means that
only algorithms with exponentially increasing running time are known.
However, for the special case of a planar system without magnetic field, e.g.
a square lattice with periodic boundary conditions in at most one direction,
there are
efficient polynomial-time ``matching'' algorithms\cite{bieche1980}. The
basic idea is to represent each realization of the disorder by its frustrated
plaquettes\cite{toulouse1977}. Pairs of
frustrated plaquettes are connected by paths
in the lattice and the weight of a path is defined by the sum of the absolute
values of the coupling constants which are crossed by the path.
A ground state corresponds the set of paths with minimum total
weight, such that each frustrated plaquette is connected to exactly one
other frustrated plaquette. This is called a minimum-weight perfect
matching.
The bonds which are crossed by paths
connecting the frustrated plaquettes are unsatisfied in 
the ground state, and all other bonds are satisfied.

For the calculation of the minimum-weight perfect matching, efficient
polynomial-time algorithms are available\cite{barahona1982b,derigs1991}.
Recently, an implementation has been presented\cite{palmer1999}, where 
ground-state energies of large systems of size $N\le 1800$ were calculated. 
Here, an algorithm from the LEDA library\cite{leda1999} has been applied,
which limits the system sizes to $N\le 480^2$, due to the restricted
size of the main memory of the computers which we used.

\begin{figure}
\begin{center}
\epsfxsize=0.93\columnwidth
\epsfbox{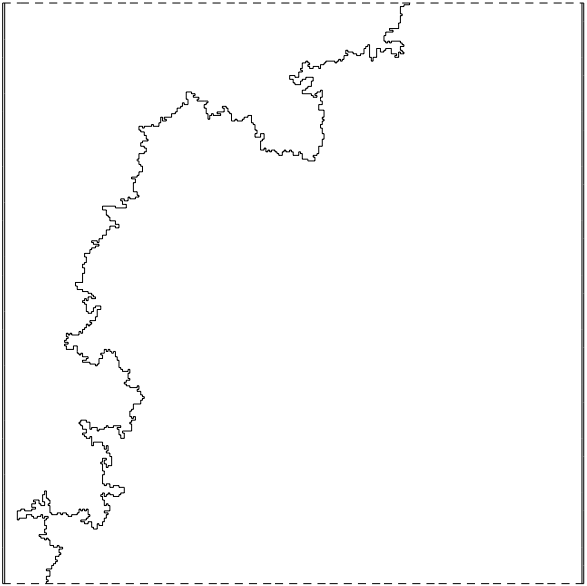}
\end{center}
\caption{A domain wall created in an $L=320$ system with Gaussian interactions.
Free boundary conditions are applied across the top and bottom edges (dashed
lines) and periodic boundary conditions applied across the vertical edges
(solid lines). The domain wall is induced by changing the periodic boundary
conditions to
antiperiodic, which is equivalent to changing the sign of
the bonds which wrap round the system from left to right.
}
\label{wall}
\end{figure}

To study whether an ordered phase is stable at finite temperatures,
the following procedure is usually applied 
\cite{mcmillan1984,bray84,rieger1996,kawashima1997,%
alex-stiff,alex-4d,alex-rcf}.
First a ground state 
of the system is calculated, having energy $E_0$.
Then the system is perturbed to introduce a domain wall and the
new ground state energy, $E_0^{\rm pertb}$ is evaluated. Typically, the
system initially has periodic boundary conditions in both directions, and the
perturbation involves replacing periodic by anti-periodic boundary
conditions in one direction. However we cannot apply the matching algorithm
for boundary conditions which ``wrap around'' in both directions, so
instead we have investigated two slightly different procedures:
\begin{enumerate}
\item[(i)]
We choose boundary conditions which are
free in one direction and periodic in the other
($x$-say).
The periodic boundary conditions are then replaced by anti-periodic, which is
equivalent to changing sign of the bonds which ``wrap around'' the system. This is
very similar to the usual approach, except that the induced domain wall no
longer has to have the same $x$ coordinate at the top and bottom of the
sample. An example of a domain wall formed in this way is shown in
Fig.~\ref{wall}. We denote these as ``P-AP'' boundary condition 
changes.
Note that the
energy change can have either sign, and the distribution of values
(obtained by repeating the calculation for many samples) 
is symmetric.

\item[(ii)]
We apply free boundary conditions in both directions and compute the ground
state. We then add extra bonds which wrap around the system in the
$x$-direction say, and have a sign and strength such that they force the spins
they connect to have the opposite relative orientation to that which
they had in the original ground state. The new ground state is then computed
and we remove from the ground state energy the 
contribution from the added
strong bonds. The change in energy must then be positive. 
This procedure is similar to
the ``conjugate boundary conditions''
chosen in Ref.~\onlinecite{matsubara1998}. 
Since free boundary conditions are still applied along the top and
bottom edges, the domain walls look very similar to the one shown in
Fig.~\ref{wall}.  We denote this boundary condition change by
``F-DW'', since the
initial boundary condition is {\bf F}ree, and it is then changed to induce a
{\bf D}omain {\bf W}all.

\end{enumerate}

For a given sample, the domain-wall energy is given by 
\begin{equation}
\delta E=E_0^{\rm pertb}-E_0  \, .
\end{equation}
To study finite-size behavior we consider the ``defect energy'', $\Delta E$,
defined by
\begin{equation}
\Delta E = [\, | \delta E |\, ]_{\rm av} ,
\label{eq:de}
\end{equation}
where $[ \cdots ]_{\rm av}$ denotes an average over samples. (Of course the
absolute value is not necessary for the F-DW boundary conditions since $\delta
E \ge 0$.)
It is expected that $\Delta E$ will vary as
\begin{equation}
\Delta E \sim L^{\theta} ,
\label{eq:powerlaw}
\end{equation}
where $\theta$ is a stiffness exponent.
For 
$\theta<0$ an ordered phase is only stable at $T=0$ and the correlation
length $\xi$ diverges\cite{bray84} 
for $T \to 0$ as $\xi \sim T^{1/\theta}$. 
For $\theta>0$
an ordered phase also occurs for $T>0$. At the lower critical dimension,
one has
$\theta = 0$ and expects an exponential divergence of the correlation
length.

Domain-wall energies obtained in the way described above were obtained
for systems with Gaussian and $\pm J$ disorder, for both F-DW and P-AP
boundary condition changes.
Sizes up to $N=480^2$ were considered, and, for each
size,
the result was average over up to $40000$ samples. Table \ref{params} shows the number
of samples for each system size and type of system.

\begin{table}[ht]
\begin{tabular}{l|rrrr}
type & $L\le 160$ & $L=240$ & $L=320$ & $L=480$ \\\hline
$\pm J$ P-AP & 40000 & 45000 & 24000 & 5000 \\
$\pm J$ F-DW & 10000 & 5500 & 4500 & -\\
Gaussian P-AP & 10000 & 2800 & 3700 & -\\
Gaussian F-DW & 10000 & 700   & -  & -
\end{tabular}
\caption{The number of samples for each type of disorder and system size.}
\label{params}
\end{table}

\begin{figure}[htb]
\begin{center}
\epsfxsize=0.93\columnwidth
\epsfbox{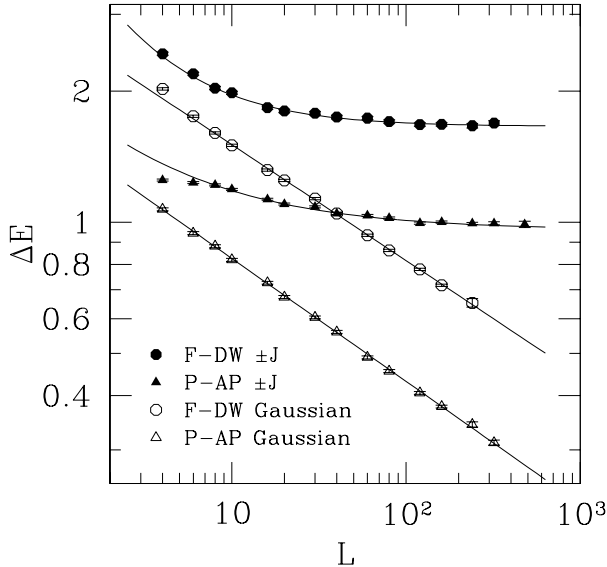}
\end{center}
\caption{The defect energy $\Delta E$ as a function of system size
for the different types of disorder and boundary condition changes.
The lines are fits (for $L\ge 8$)
to the form $\Delta E = \Delta E_{\infty}+bL^{\theta}$. The following
fit parameters were obtained: 
$\Delta E=0.96$, $b=0.99$, $\theta=-0.65$ ($\pm J$ P-AP);
$\Delta E=1.66$, $b=2.93$, $\theta=-0.99$ ($\pm J$ F-DW);
$b=1.58$, $\theta=-0.282$ ($\Delta E\equiv 0$, Gaussian P-AP);
$b=2.78$, $\theta=-0.266$ ($\Delta E\equiv 0$, Gaussian F-DW).
}
\label{figStiff}
\end{figure}

In Fig. \ref{figStiff} the defect energies are shown
as a function of system size. We first discuss the results for
the Gaussian distribution.
The domain-wall energy clearly decreases with a power law 
$\Delta E \sim L^{\theta}$ with 
\begin{eqnarray}
\theta & = & -0.282 \pm 0.002 \qquad {\rm (P-AP)} \nonumber \\
\theta & = & -0.266 \pm 0.002 \qquad {\rm (F-DW)} .
\label{eq:theta}
\end{eqnarray}
For the goodness of the fits\cite{goodness} we obtain
$Q=0.48$ and $Q=0.62$  for the P-AP and F-DW boundary conditions
respectively.
In addition, we tried fits of the form 
$\Delta E =\Delta E_{\infty}+bL^{\theta}$, obtaining
$\Delta E_{\infty}=0.04 \pm 0.02$ 
($Q=0.48$) for P-AP boundary conditions
and $\Delta E_{\infty}=-0.03 \pm 0.07$ ($Q=0.54$) for
F-DW boundary conditions.
Hence, we cannot rule out that the domain-wall
energy converges to a small nonzero value for $L \to \infty$
but this seems unlikely.

The result of the exponent for the P-AP boundary conditions
is compatible with the value of $\theta=-0.281(2)$ found\cite{rieger1996}
for full periodic boundary conditions 
and system sizes $L\le 30$. The value of  the F-DW case disagrees with 
the value for the P-AP case by more than the error bars, but
the difference is quite small. We expect that this
is due to systematic corrections to scaling and that asymptotically
the same result would be obtained, i.e. the same exponent would be
obtained for any reasonable definition of the defect energy.
Our result for the F-DW boundary condition differs significantly from the
value
$\theta=-0.20$ obtained\cite{matsubara1998}.
The reason for the
discrepancy may be that Ref. \onlinecite{matsubara1998}
 studied smaller sizes, $L \le 24$, and
used a Monte-Carlo method to determine the ground states instead
of an exact algorithm. Nevertheless, to our knowledge, 
all results obtained for the Gaussian model are compatible with
$T_c=0$.

Next we turn to the results for the
$\pm J$ model. The difference between these results and those
for the
Gaussian model, both of which are shown in Fig.~\ref{figStiff},
is quite striking.
Whereas the data for the Gaussian
model clearly tends to zero with a
power law for $L \to \infty$, the data for the $\pm J$ distribution appears to
tend to a non-zero value. Fits of the same form as for the Gaussian
model give
$\Delta E_{\infty}=0.96 \pm 0.01$ ($Q=0.23$) for the P-AP boundary conditions
and
$\Delta E_{\infty}=1.66 \pm 0.01$ ($Q=0.50$) for the
F-DW boundary conditions. Fits in which 
$\Delta E_{\infty}$ was fixed to zero were also
tried but gave
ridiculously low probabilities: $Q=2\times 10^{-27}$ and $Q=2\times 10^{-17}$
respectively.
For the case with free boundary conditions in the x-direction, a
saturation has been observed before\cite{matsubara1998} for small system
sizes $L\le 24$, using a Monte-Carlo method to find ground states, rather
than an exact algorithm.

\begin{figure}[htb]
\begin{center}
\epsfxsize=0.93\columnwidth
\epsfbox{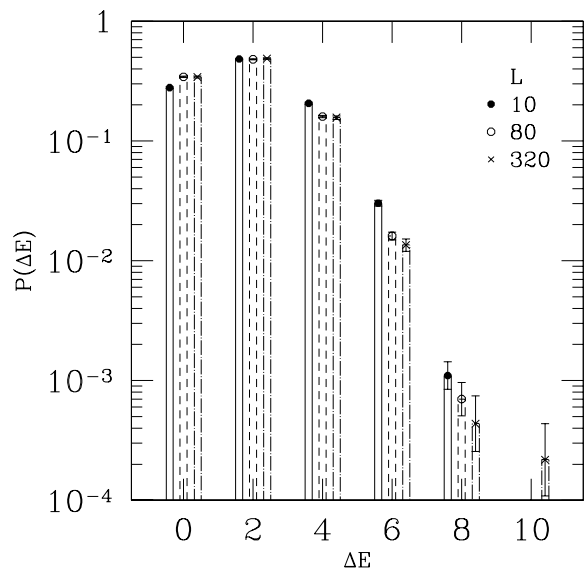}
\end{center}
\caption{Distribution $P(\Delta E)$ of domain-wall energies for the
$\pm J$ distribution with F-DW boundary conditions. The points for
$L=320$ and $L=10$ are slightly displaced on the x-axis for visibility.
}
\label{figPEpm}
\end{figure}

\begin{figure}[htb]
\begin{center}
\epsfxsize=0.93\columnwidth
\epsfbox{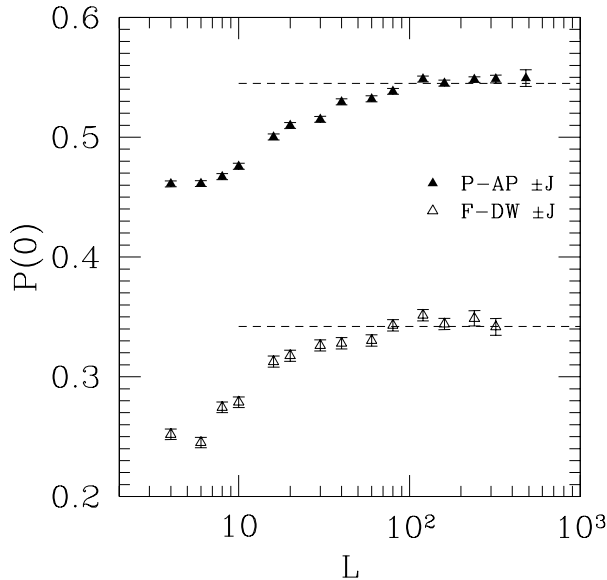}
\end{center}
\caption{The probability $P(0)$ that the domain-wall
energy is exactly zero for the $\pm J$ model as a function of system
size. The lines are guides to the eye.}
\label{figP0}
\end{figure}

To show the effect of saturation into more detail, Fig.~\ref{figPEpm}
displays the distribution of the domain-wall energies, with a logarithmic
vertical scale, for the $\pm J$ distribution with F-DW boundary conditions.
For this distribution, only even integer values
of $\Delta E$ are found.
Except for the largest values of $\Delta E$, where the weights are
very small, the weights of the discrete bins are almost the same for $L=80$ and $320$
indicating that the whole distribution
converges for large $L$. 
The probability $P(0)$ that the domain-wall
energy is exactly zero is displayed in Fig.~\ref{figP0}. 
$P(0)$ seems to converge towards a finite value less than unity
for both boundary conditions.

From this data we deduce that 
$\theta = 0$ for the $\pm J$ distribution, so
the lower critical dimension is $d_c=2$ exactly. The most likely
scenario at finite temperature is that the correlation length diverges
exponentially as $T \to 0$, perhaps like $\exp(C/T)$ or $\exp(C/T^2)$.
Detailed Monte Carlo simulations\cite{houdayer2001}
and the work of Saul and
Kardar\cite{saul1993} indicate that the former
is the case.
The other possibility is a finite-$T$ Kosterlitz-Thouless
type of transition without true long range order below $T_c$, as claimed by
Ref.~\onlinecite{shirakura1997}. However, the finite value for $\nu$ obtained
in Ref.~\onlinecite{shirakura1997} seems incompatible with this picture.
The reason for this discrepancy is probably that the system sizes
were  quite small, $N\le 18^2$.

To conclude, we have studied
the defect energy of the
two-dimensional
Ising spin glass
using
much larger system sizes than before, up to
$N=480^2$.
For Gaussian disorder
we find that a spin-glass phase is not stable at $T>0$, and that $d_c > 2$,
in agreement with earlier findings for much smaller systems.
For the $\pm J$
model, the absolute value of the domain-wall 
energy saturates at a nonzero value for $L\to \infty$,
indicating that the lower critical dimension is exactly $d_c=2$, in agreement
with  Refs.~\onlinecite{houdayer2001} and \onlinecite{saul1993}.
It is quite striking and perhaps
surprising that the $\pm J$ model, which has a large ground state degeneracy,
actually has {\em more}\/ order at very low-$T$ than the Gaussian
distribution, which has a unique ground
state.

\acknowledgements

The simulations were performed at the Paderborn Center
for Parallel Computing (Germany) and on a Beowulf Cluster
at the Institut f\"ur Theoretische Physik of the Universit\"at
Magdeburg (Germany). AKH acknowledges financial support from the DFG (Deutsche 
Forschungsgemeinschaft)
under grant Ha 3169/1-1. APY acknowledges support
from the NSF through grant DMR 0086287.


\begin{thebibliography}{99}

\bibitem[*]{alex:email} hartmann@bach.ucsc.edu

\bibitem[\dagger]{peter:email} peter@bartok.ucsc.edu

\bibitem{revievSG}  
    K. Binder and A.P. Young, Rev. Mod. Phys. {\bf 58}, 801 (1986); M. Mezard,
    G. Parisi, M.A. Virasoro, {\it Spin glass theory and beyond}\/, (World
    Scientific, Singapore 1987); K.H. Fisher and J.A. Hertz, {\em Spin
    Glasses}\/, (Cambridge University Press, Cambridge 1991); A.P. Young
    (ed.), {\em Spin glasses and random fields}\/, (World Scientific,
    Singapore 1998).

\bibitem{mcmillan1984}
    W.L. McMillan, Phys. Rev. B {\bf 30}, 476 (1984).

\bibitem{bray84}
    A.J. Bray and M.A. Moore, J. Phys. C {\bf 17}, L463 (1984);
    A.J. Bray and M.A. Moore in
    {\em Heidelberg Colloquium on Glassy Dynamics}\/,
    J.L. van Hemmen and I. Morgenstern (eds),
    (Springer-Verlag, Heidelberg 1987).

\bibitem{rieger1996}
    H. Rieger, L. Santen, U. Blasum-U, M. Diehl, M. J\"unger, and G. Rinaldi, J.
    Phys. A {\bf 29}, 3939-50 (1996).

\bibitem{kawashima1997}
    N. Kawashima and H. Rieger, Europhys. Lett. {\bf 39}, 85 (1997).

\bibitem{shirakura1997}
    T. Shirakura and F. Matsubara, Phys. Rev. Lett.  {\bf 79}, 2887 (1997).

\bibitem{matsubara1998}
    F. Matsubara, T. Shirakura, and M. Shiomi, Phys. Rev. B {\bf 58}, R11821
    (1998).

\bibitem{houdayer2001}
    J. Houdayer, preprint cond-mat/0101116.

\bibitem{saul1993}
    L. Saul and M. Kardar, Phys. Rev. E {\bf 48}, R3221 (1993). 
    This work uses
    a polynomial-time algorithm which works with periodic boundary conditions
    in both directions (and even for $T > 0$) but is less efficient than the
    matching algorithm used here. 

\bibitem{barahona1982}
    F. Barahona, J. Phys. A {\bf 15}, 3241 (1982).

\bibitem{opt-phys2001}
    A.K. Hartmann and H. Rieger, {\em Optimization Algorithms in Physics}\/,
    Wiley-VCH, Berlin, to be published 2001.

\bibitem{bieche1980}
    I. Bieche, R. Maynard, R. Rammal, and J.P. Uhry, J. Phys. A {\bf 13}, 2553
    (1980).

\bibitem{toulouse1977} G. Toulouse, Commun. Phys. {\bf 2}, 115 (1977).

\bibitem{barahona1982b}
    F. Barahona, R. Maynard, R. Rammal, and J.P. Uhry, J. Phys. A {\bf 15},
    673 (1982).

\bibitem{derigs1991}
    U. Derigs and A. Metz, Math. Prog. {\bf 50}, 113 (1991).

\bibitem{palmer1999}
    R.G. Palmer and J. Adler, Int. J. Mod. Phys. C {\bf 10},  667 (1999).

\bibitem{leda1999} 
    K. Mehlhorn and St. N\"aher, {\em The LEDA Platform of Combinatorial and
    Geometric Computing}\/, Cambridge University Press, Cambridge 1999; see
    also  {\tt http://www.mpi-sb.mpg.de/LEDA/leda.html}.

\bibitem{alex-stiff}
    A.K. Hartmann, Phys. Rev. E {\bf 59}, 84 (1999).

\bibitem{alex-4d}
    A.K. Hartmann, Phys. Rev. E {\bf 60}, 5135 (1999).

\bibitem{alex-rcf}
    A.K. Hartmann and I.A. Campbell, Phys. Rev. B {\bf 63}, 094423 (2001).

\bibitem{goodness}
    $Q$ is the probability that the value of $\chi^2$
    is worse than in the
    current fit. See  W.H. Press, S.A. Teukolsky, W.T. Vetterling and B.P.
    Flannery, {\em Numerical Recipes in C}\/, Cambridge University Press,
    Cambridge (1995).


\end{thebibliography}
\end{document}